\documentclass[a4paper,11pt,hyper]{JHEP3}
\usepackage{amsfonts,latexsym,graphicx,epsfig,amssymb,amsmath,mathrsfs, bmpsize}
\usepackage{ulem}

\newcommand{\dd}{{\rm d}}

\newcommand{\cR}{{\cal R}}
\newcommand{\bm}[1]{\hbox{\boldmath{$#1$}}}
\newcommand{\sbm}[1]{\hbox{\boldmath{\scriptsize$#1$}}}

\newcommand{\bw}{\begin{widetext}}
\newcommand{\ew}{\end{widetext}}%
\newcommand{\be}{\begin{equation}}
\newcommand{\ee}{\end{equation}}
\newcommand{\beqa}{\begin{eqnarray}}
\newcommand{\eeqa}{\end{eqnarray}}

\newcommand{\D}{\Delta}
\renewcommand{\a}{\alpha}
\renewcommand{\b}{\beta}
\newcommand{\g}{\gamma}

\newcommand{\vk}{\varkappa}
\newcommand{\ve}{\varepsilon}

\renewcommand{\ln}{\mathop{\rm ln}\nolimits}

\title{Inflationary perturbations with Lifshitz scaling
}
\author{
Shun Arai$^a$,\, Sergey Sibiryakov$^{b,c,d}$,\, Yuko Urakawa$^{a,e}$\\
a.\,Department of Physics and Astrophysics, Nagoya University, Chikusa,
Nagoya 464-8602, Japan\\
b.\,Theoretical Physics Department, CERN, CH-1211 Gen\`eve 23,
Switzerland\\
c.\,FSB/IPHYS/LPPC, \'Ecole Polytechnique F\'ed\'erale de Lausanne,
\normalsize\it CH-1015, Lausanne, Switzerland \\
d.\,Institute for Nuclear Research of the
Russian Academy of Sciences, \\ 
\normalsize \it  60th October Anniversary Prospect, 7a, 117312
Moscow, Russia\\
e.\, Institut de Ciencies del Cosmos,
Universitat de Barcelona, Marti i Franques 1, 08028 Barcelona, Spain
}

\abstract{Instead of Lorentz invariance, gravitational degrees of
  freedom may obey Lifshitz scaling at high energies, as it happens in
  Ho\v rava's proposal for quantum gravity. We study consequences of
  this proposal for the spectra of primordial perturbations generated
  at inflation. Breaking of 4D diffeomorphism (Diff) invariance down
  to the foliation-preserving Diff in Ho\v{r}ava-Lifshitz (HL) gravity leads to
  appearance of a scalar degree of freedom in the gravity sector, khronon, which
describes dynamics of the time foliation. One can
naively expect that mixing between inflaton and khronon will
jeopardize conservation of adiabatic perturbations at super
Hubble scales. This indeed happens in the projectable
version of the theory.
By contrast, we
find that in the non-projectable version of HL gravity, khronon 
acquires an effective mass which is much larger than the
Hubble scale well before the Hubble crossing time and decouples
from the adiabatic 
curvature perturbation $\zeta$ sourced by the inflaton 
fluctuations. 
As a result,  at super Hubble scales the adiabatic perturbation $\zeta$
behaves as in an effectively single field system and its spectrum is
conserved in time. Lifshitz scaling is imprinted in the power
spectrum of $\zeta$ through the modified dispersion
relation of the inflaton. We point out violation of the consistency relation
between the tensor-to-scalar ratio and the spectral tilt of primordial
gravitational waves and suggest that it can provide a signal of
Lorentz violation in 
inflationary era.}

\keywords{Lorentz violation, Inflation, 
Lifshitz theories}
\preprint{CERN-TH-2018-040, INR-TH-2018-001}

\begin{document}



\section{Introduction}\label {Sec:intro}

General relativity (GR) accurately describes all known gravitational
phenomena. Still, it has a theoretical flaw: it is not
renormalizable~\cite{Goroff_Sagnotti1986} and thus cannot be a
complete theory of quantum gravity. 
One way to address this problem is to
introduce terms with higher powers of the curvature tensor which make the theory
renormalizable~\cite{Stelle1977}. However, if Lorentz invariant, these
higher curvature terms lead to loss of unitarity. This motivated
P. Ho\v{r}ava to propose a framework to render gravity power-counting
renormalizable by
abandoning Lorentz invariance~\cite{Horava:2009uw}.
By breaking Lorentz invariance, we can introduce higher spatial
derivative terms, while avoiding higher time derivative terms and
thus making the theory compatible with unitarity.
A key role in the power-counting argument is played by an approximate
invariance of the theory at high energies and momenta with respect to
the so-called Lifshitz scaling transformations. These stretch space
and time by different amount, so they are also often referred to as
anisotropic scaling. The dispersion relations of various degrees of
freedom at high energies, compatible with anisotropic scaling, have
the form $\omega\propto p^z/M_*^{z-1}$, where $\omega$ and $p$ are
particle's energy and momentum, $z$ is the Lifshitz exponent ($z$
equals the number of spatial dimensions in Ho\v rava's proposal)
and $M_*$ is the energy threshold, above which the
anisotropic scaling sets in.   
This framework has received the name of Ho\v{r}ava--Lifshitz (HL)
gravity and the so-called projectable subclass of the resulting
theories has been rigorously demonstrated to be perturbatively
renormalizable~\cite{Barvinsky:2015kil,Barvinsky:2017zlx}. Moreover,
in 2 spatial and 1 time dimensions the theory 
exhibits asymptotic freedom~\cite{Barvinsky:2017kob} which strongly
suggests that it is ultraviolet (UV) complete. 

Deviations from Lorentz invariance are tightly constrained in the
Standard Model sector~\cite{Mattingly:2005re,Kostelecky:2008ts,
Liberati:2013xla}. In the gravity sector constraints come from observations at low
energies such as Solar
System tests~\cite{Liberati:2013xla, Will:2005va}, pulsar
timing~\cite{Yagi:2013qpa,Yagi:2013ava,  Shao:2013wga, Jimenez:2015bwa}, 
cosmology~\cite{Saridakis:2009bv,Dutta:2009jn,Kobayashi:2010eh,APSG,
Blas:2012vn,Audren:2013dwa,Audren:2014hza} 
and direct detection of the gravitational waves~\cite{Blas:2016qmn,
  Yunes:2016jcc,Monitor:2017mdv,Gumrukcuoglu:2017ijh}. 
By contrast, Lorentz violation (LV) in the gravity sector is poorly
constrained at high energies 
where it is motivated by renormalization
of gravity.

In order to examine the consequences of LV at high energies we
study in this paper its effect on 
cosmic inflation in the early universe. 
One may expect that breaking of Lorentz invariance during inflation 
will leave an imprint on the
primordial perturbations generated during inflation. 
This possibility has been explored in a number of 
works~\cite{ArkaniHamed:2003uz,TakahashiSoda,
Calcagni:2009ar,Kiritsis:2009sh,Mukohyama:2009gg,Kobayashi:2009hh,
Donnelly:2010cr,Creminelli:2012xb,Solomon:2013iza,Ivanov:2014yla}. 
In HL gravity, where 4D diffeomorphism (Diff) is reduced to foliation
preserving Diff, there appears a scalar degree of 
freedom in gravity sector, so called khronon. 
It is tempting to speculate that this additional degree of freedom can
play the role of inflaton. However, at the moment this seems to be
forbidden due the restrictive symmetry structure of the theory. 
Therefore, to drive inflation, we
need to introduce a scalar field, 
as usual. Then, in general, generation of 
the primordial scalar perturbation is described by a coupled system
for two fields, the inflaton and khronon perturbations. 

To provide a prediction of the observable quantities, we need to
solve consistently the two field system of the inflaton and khronon,
which are coupled with each other during inflation. When 4D Diff is
preserved and the universe is dominated by a single component, it
is well-known that the adiabatic curvature perturbation
$\zeta$ stays constant in
time after the Hubble crossing (see, e.g.,
Refs.~\cite{Wands:2000dp, Weinberg:2003sw}). On the other hand, in HL
gravity the number of scalar degrees of freedom is
always greater than one due to 
the presence of khronon and it is not clear a priori
if there exists a conserved variable or not. 
 
The inflaton and khronon are gravitationally coupled even in the
absence of a direct interaction between them. In this paper we compute
the primordial power spectra by consistently solving the two field
models with the inflaton and 
khronon. The previous studies mostly focused on the regime where the
Hubble scale of inflation is low, $H<M_*$, so that the higher derivative terms
in the action are unimportant and the theory is described by its
infrared (IR) limit. By contrast, in this paper we are interested in
the high-energy regime of Lifshitz scaling relevant for the
case\footnote{Recent observation of gravitational waves from neutron
  star merger in coincidence with the electromagnetic signal
  \cite{Monitor:2017mdv} points towards an upper bound on the scale $M_*$ in
non-projectable HL gravity, $M_*\lesssim 10^{11}$GeV
\cite{Gumrukcuoglu:2017ijh}. Hence, in this theory
the Lifshitz regime is relevant whenever the
inflationary Hubble exceeds $10^{11}$GeV.
} 
$H>M_*$. 
We consider both projectable and non-projectable versions of
HL gravity. 
As discussed in Refs.~\cite{Blas:2009yd,Blas:2010hb}, 
the khronon sector of the projectable HL gravity
suffers from either the gradient instability or the strong coupling in the
IR limit. This means that it cannot describe the physics all the way
down to low energies, unless inflationary epoch is separated from the
later hot universe by a phase transition that eliminates khronon from
the spectrum. Still,
the projectable version is perfectly well-behaved in the high-energy
regime and its study is instructive to make comparison with 
the non-projectable version.

When the fluctuations are deep inside the Hubble scale, the gravitational
interaction is suppressed and we simply have two decoupled Lifshitz
scalars. On the other hand, in the super Hubble scales, the gravitational
interaction makes the inflaton and khronon coupled. Then one may naively
expect that the primordial spectrum will depend on the time evolution of
these two fields and we will need to solve the evolution all along also
after the Hubble crossing time. Indeed, this is the case for the
projectable version. On the other hand, in the non-projectable version,
we will find that khronon gets decoupled from the adiabatic curvature
perturbation $\zeta$. As a result, $\zeta$ is conserved at large scales
and the power spectrum of $\zeta$ is solely determined by the
inflaton. Thanks to the presence of the conserved quantity, we can
easily calculate the spectrum of the fluctuation at the end of
inflation. Then the consequence of the LV in the spectrum of $\zeta$
only stems from the modification of the dispersion relation. 

The spectrum of primordial gravitational waves in HL gravity was
computed in Ref.~\cite{TakahashiSoda}. Once the scalar perturbation is
obtained, we can also compute the tensor to scalar ratio $r$. In a 4D
Diff invariant theory, there exists a universal relation between $r$ and
the tensor spectral tilt $n_t$, the so-called consistency relation. We
will show that this consistency relation can be broken if the primordial
perturbations are generated in the anisotropic scaling regime. The violation of
the consistency relation provides a signal of LV
in the gravity sector in the high energy regime. 

This paper is organized as follows. In Sec.~\ref{Sec:Lif} we
describe our setup and review the computation of the power
spectrum of the Lifshitz scalar and the gravitational waves generated in
the anisotropic scaling regime. In Sec.~\ref{Sec:Khronon} we discuss the
behaviour of the khronon perturbation. We show that khronon stays
gapless in the projectable version, while it is gapped in the
non-projectable version, which leads to the decoupling from the
adiabatic mode. In Sec.~\ref{Sec:LV} we discuss violation of
the consistency relation by inflationary perturbations with Lifshitz scaling.
We conclude in Sec.~\ref{Sec:concl}.   
Appendices summarize some technical details.

\section{Primordial perturbations with anisotropic scaling}
\label{Sec:Lif}
In this section we describe our setup and briefly summarize the
computation of the primordial 
spectra of the Lifshitz scalar and gravitational
waves.  

\subsection{Projectable and non-projectable Ho\v{r}ava gravity}
\label{ssec:HL}
\subsubsection{Lagrangian densities}
First, we consider the non-projectable version of HL gravity
\cite{Horava:2009uw} with the extension introduced in
\cite{Blas:2009qj}. Due to the complexity of the most general Lagrangian in this framework, we restrict only to
the terms that contribute to the action at quadratic order in the
perturbations around spatially flat backgrounds and that preserve the
parity invariance. This restriction is sufficient to capture the
qualitative features of the theory.
The complete list of
these terms is given in \cite{Blas:2009qj} and leads to the following
Lagrangian density,
\begin{align}
 {\cal L}_{HG} =N\sqrt{h}\bigg\{&\frac{M_*^2}{2}
\bigg[\frac{1}{\a_1}K_{ij}K^{ij}-\frac{1}{\a_2}K^2+\frac{1}{\a_3}R+a_i
a^i\bigg] \cr
&  -\frac{1}{2}\bigg[\frac{R_{ij}R^{ij}}{\b_1}+\frac{R^2}{\b_2}
-\frac{R\nabla_ia^i}{\b_3}+\frac{a_i\D a^i}{\b_4}\bigg] \cr
&  -\frac{1}{2M_*^2}\bigg[\frac{(\nabla_iR_{jk})^2}{\gamma_1}
+\frac{(\nabla_i R)^2}{\g_2}+\frac{\D R\nabla_ia^i}{\g_3}
-\frac{a_i\D^2 a^i}{\g_4}\bigg]\bigg\}, 
\label{Lagrfull}
\end{align}
where we used the ADM line element, given by
\begin{equation}
 \dd s^2=(N^2-N_iN^i)\dd t^2-2N_i\dd t \dd x^i-h_{ij}\dd x^i \dd x^j\,.
\end{equation}
Here $R_{ij}$, $\nabla_i$ and $\Delta$ denote the 3-dimensional Ricci
tensor, 
the covariant derivative with respect to $h_{ij}$ and the covariant
Laplacian,
\begin{equation}
K_{ij}=\frac{\dot h_{ij}-\nabla_i N_j-\nabla_j N_i}{2N}
\end{equation}
is the extrinsic curvature and we have defined $a_i$ as 
\begin{align}
  a_i \equiv \frac{ \partial_iN}{N}\,.
\end{align}
Note that we included the integration measure in the definition of the
Lagrangian density. The terms in the first line of
Eq.~(\ref{Lagrfull}) describe the low energy part of the action, and
the 
parameters entering it are constrained by the present-day 
observations\footnote{We assume that during inflation these parameters
  have the same values as nowadays. This assumption can
  be relaxed in a more general setup.}. The relation between
these parameters and 
the parameters $\a,\lambda,\xi$
introduced in \cite{Blas:2010hb} is
\begin{align}
 & \label{relpar}
M_*^2=M_P^2\a\;,~~ \a_1=\a\;,~~\a_2=\a/\lambda\;,~~\a_3=\a/\xi\;,
\end{align} 
where $M_P$ is the Planck mass. In what follows, we will write 
\begin{align}
 & \alpha_1 - \alpha_2  = 2 \alpha_1 \bar{\alpha} \,.
\end{align}

We also discuss the projectable version, where the lapse function is
postulated to be space-independent,
\begin{align}
 & N = N(t)\,.
\end{align} 
The action for the projectable version can be obtained simply by dropping the
perturbation of the lapse function in the action for the non-projectable
version. Then the parameters $\beta_3$, $\beta_4$, $\gamma_3$, and
$\gamma_4$ are irrelevant in the projectable theory. 

For both the non-projectable and projectable versions, we add as the inflaton
a Lifshitz scalar field whose Lagrangian density is given by: 
\begin{align}
 & \label{LagrInf}
{\cal L}_{inf}=N\sqrt{h}\biggl\{\frac{(\dot\Phi-N^i
 \partial_i\Phi)^2}{2N^2}
-   \frac{\vk_1}{2}  \nabla_i \Phi \nabla^{i} \Phi -
  \frac{\vk_2}{2 M_*^2}  \nabla_i\nabla_j  \Phi \nabla^i\nabla^j \Phi \cr
 & \qquad \qquad   \qquad \qquad   
 - \frac{\vk_3}{2M_*^4}  \nabla_i\nabla_j \nabla_k \Phi \nabla^i \nabla^j\nabla^k  \Phi-V(\Phi)\biggr\}\;.
\end{align} 
In principle, the coefficients $\vk_{1,2,3}$ here can be functions of
the field $\Phi$ which has zero scaling dimension. We concentrate on
the case of constant coefficients for simplicity.
We assume that the inflaton is minimally coupled to the gravity
sector. We will briefly discuss a non-minimally coupled case in
Sec.~\ref{Sec:concl}.

\subsubsection{Parameter hierarchy}
The Lagrangian density (\ref{Lagrfull}) contains a number of
parameters. Here we discuss the hierarchy between them. Stability and
constraints on deviations from Lorentz invariance 
at low energies require~\cite{Blas:2010hb},
\begin{align}
 & 0 < \alpha_1 \ll 1\,.
\end{align}
Consider now 
the propagation of gravitational waves in flat spacetime where their
dispersion relation is given by
\begin{align}
 & \omega^2(p) = p^2\,
 \sum_{z=1}^3 \vk_{\gamma,z} \left( \frac{p}{M_*} \right)^{2(z-1)} 
\, ,
\end{align}
with
\begin{align}
 & \vk_{\gamma,1} \equiv  \frac{\alpha_1}{\alpha_3}\,, \qquad  \vk_{\gamma,2} \equiv
 \frac{\alpha_1}{\beta_1}\,, \qquad  \vk_{\gamma,3} \equiv \frac{\alpha_1}{
 \gamma_1}\,. \label{Def:vk}
\end{align}
The coefficient $\vk_{\gamma, 1}$ determines (the square
of) the propagation speed of the gravitational waves at low energies. 
According to the constraints from the observation of the Hulse-Taylor
pulsar~\cite{Jimenez:2015bwa} and more directly from the detections of
the gravitational waves at the two detector sites~\cite{Blas:2016qmn}, 
the propagation speed of the gravitational
waves in the IR should be of order of the speed of light, which imposes 
$\alpha_1 \simeq \alpha_3$.  The recent detections of GW170817 and
GRB170817A give a tight constraint 
$|\vk_{\gamma,1}-1|<10^{-15}$ \cite{Monitor:2017mdv}. (See also
Ref.~\cite{Moore:2001bv} for the constraint on the subluminal
propagation of the gravitational waves from the absence of
the gravitational Cherenkov radiation.) 
Next, requiring that the transition from linear dispersion relation to
the Lifshitz scaling happens at $p\sim M_*$ we obtain the requirements
$\vk_{\g,2}, \vk_{\g,3}\simeq 1$. 
By combining these
two conditions, we obtain
\begin{align}
 & \alpha_1 \simeq  \alpha_3 \simeq \beta_1 \simeq \gamma_1 \ll 1
 \,. \label{parameterGW} 
\end{align}

Let us now turn to khronon.
In the projectable version its dispersion relation reads,
\begin{align}
 & \omega_{pr}^2(p) =  \frac{\alpha_1\bar{\alpha}
 }{1+ \bar{\alpha}} p^2  \left[ -\frac{1}{\alpha_3} + \biggl( \frac{3}{\beta_1} + \frac{8}{\beta_2}
 \biggr)  \biggl( \frac{p}{M_*} \biggr)^2+ \biggl( \frac{3}{\gamma_1} + \frac{8}{\gamma_2}
 \biggr)  \biggl( \frac{p}{M_*} \biggr)^4 \right]\,. \label{Exp:DRP}
\end{align}
The first term in the square brackets is negative and is responsible
for gradient instability in IR. On the other hand, the remaining
terms in (\ref{Exp:DRP}) can be chosen positive, so that at $p>M_*$
the dispersion relation is well-behaved.
Again,
setting the transition to Lifshitz scaling 
at around $p \simeq M_*$ and taking into account (\ref{parameterGW}) we
obtain 
\begin{align}
 & \alpha_{1,3} \simeq \beta_{1,2} \simeq \gamma_{1,2} \ll 1\,. \label{parameterP}
\end{align}
Further requiring that the overall magnitude of the 
frequency $\omega_{pr}(p)$
in UV is 
${\cal O}(p^z/M_*^{z-1})$ we set $\bar{\alpha} \simeq {\cal O}(1)$.
To sum up, in the projectable case we will work under the assumptions,
\begin{equation}
\label{parameterProj}
\a_{1,2,3}\simeq\b_{1,2}\simeq\g_{1,2}\ll 1\,,\qquad\bar\a={\cal O}(1)
\qquad\qquad \text{(projectable).}
\end{equation}

In the non-projectable version, the dispersion relation for khronon
becomes more complicated and is given by 
\begin{align}
 & \omega_{npr}^2(p) =  \omega_{pr}^2(p)+ \frac{2\alpha_1 \bar{\alpha}}{1+ \bar{\alpha}}\, p^2 \, \frac{\left[  -
 \frac{1}{\alpha_3} + \frac{1}{\beta_3}  (
 \frac{p}{M_*})^2 + \frac{1}{\gamma_3}  (
 \frac{p}{M_*})^4 \right]^2 }{1+ \frac{1}{\beta_4}  (
 \frac{p}{M_*})^2 + \frac{1}{\gamma_4}  (
 \frac{p}{M_*} )^4 }\;, \label{Exp:DRNP}
\end{align}
where the second piece comes from integrating out the lapse function
$N$ which enters into the action without time derivatives.
Setting the transition scale at $p \simeq M_*$ and using
Eq.~(\ref{parameterGW}), we obtain 
\begin{align}
 & \alpha_{1,3} \simeq \beta_{1,2,3} \simeq \gamma_{1,2,3} \ll 1
 \,,\qquad \beta_4 \simeq \gamma_4 = {\cal O}(1)\,. \label{parameterNP0} 
\end{align}
Similarly to the discussion of the projectable version, we assume that
$\omega(p)$ becomes ${\cal O}(p^z/M_*^{z-1})$ in UV and obtain
\begin{align}
 & \bar{\alpha} \simeq \g_3^2/\a_1  \ll 1 \,.  
\end{align}
Notice that the order of $\bar{\alpha}$ in the
non-projectable version is different from the one in the projectable
version, cf. Eq.~(\ref{parameterProj}). 
Combining all 
conditions together, we obtain 
\begin{align}
 & \alpha_{1,2,3} \simeq \beta_{1,2,3} \simeq \gamma_{1,2,3} \simeq
 \bar{\alpha} \ll 1 \,,\qquad \b_4\simeq\g_4={\cal O}(1)
\qquad\qquad \text{(non-projectable).}
\label{parameterNP} 
\end{align}
The parameters which satisfy these
conditions are consistent with the experimental data in
IR\footnote{We leave aside the question of stability of the parameter
  hierarchy under radiative corrections.}~\cite{Blas:2010hb}.

\subsection{Background equations}
\label{ssec:bg}
Equations for the inflationary background read,
\begin{align}
 &3M_P^2\frac{1+\bar\a}{1-2\bar\a} H^2
 = \frac{\dot\phi^2}{2}+V\,,  \label{FRWeqs} \\
 & \ddot\phi+ 3 H\dot\phi+V_\phi=0\,, \label{FRWeq3}
\end{align}
where $\phi$ is the background value of the inflaton and 
$V_\phi$ denotes the derivative of $V$ with respect to $\phi$.
Positivity of the l.h.s. in the Friedmann equation (\ref{FRWeqs}) 
requires $\bar{\alpha}$ to be in the range $- 1 < \bar{\alpha} < 1/2$. 
We define the slow-roll parameters,
\begin{align}
 & \varepsilon_1 \equiv - \frac{\dot{H}}{H^2}
= \frac{1- 2 \bar{\alpha}}{2(1+\bar{\alpha})} \left( 
\frac{\dot{\phi}}{M_P H} \right)^2  \,, \label{epsilon1}
\end{align}
and
\begin{align}
 & \varepsilon_n = \frac{\dd \ln \varepsilon_{n-1}}{\dd \ln a}\,,
\end{align}
 for $n\geq 2$. The expressions for the slow-roll 
parameters agree with the standard ones up to ${\cal O}(\bar{\alpha})$
 corrections. Using $\varepsilon_2$ we can express the second
derivative of $\phi$ as
\begin{align}
 & \frac{\ddot{\phi}}{H \dot{\phi}} = \frac{\varepsilon_2}{2}
 - \varepsilon_1\,. 
\end{align}
We also define the slow-roll parameters $\varepsilon_V$ and $\eta_V$ as
\begin{align}
&\varepsilon_V \equiv \frac{M^2_P}{2}\left(\frac{V_\phi}{V}\right)^2 = \frac{1-2\bar{\alpha}}{1+\bar{\alpha}} \varepsilon_1 + {\cal O}(\varepsilon^2)\,, \label{epsilonV}\\
&\eta_V \equiv M^2_P \frac{V_{\phi \phi}}{V} =
 \frac{1-2\bar{\alpha}}{1+\bar{\alpha}}\left( 2\varepsilon_1-
 \frac{\varepsilon_2}{2} \right )+{\cal O}(\varepsilon^2)\,,
 \label{etaV} 
\end{align}
where $V_{\phi \phi} \equiv \dd^2 V/\dd \phi^2$. 
In the limit $\bar{\alpha} \rightarrow 0$ the relations between
$(\varepsilon_1,\, \varepsilon_2)$ and $(\varepsilon_V,\, \eta_V)$
agree with those in GR.

\subsection{Lifshitz scalar in a fixed background}
\label{ssec:Lif}
As a warm-up exercise, in this subsection we briefly review the
computation of the spectrum of a probe massless scalar 
field $\varphi$ in a fixed inflationary
background. From now on we will work in conformal time $t$ and denote
derivatives with respect to it by primes. The action for Fourier modes
of the field reads,
\begin{align}
 & S_{scalar}  = \frac{1}{2} \int \dd t \int \dd^3 \bm{p}\,a^2 \left[
 \varphi'_{\sbm{p}} \varphi'_{- \sbm{p}} -
 {\omega^2_\varphi(t,\, p)}  \varphi_{\sbm{p}}  \varphi_{-\sbm{p}} \right]\,.
\end{align}
Anisotropic scaling in UV implies modified dispersion
relation~\cite{Mukohyama:2009gg}, 
\begin{align}
 & \frac{\omega_\varphi^2 (t, p)}{{\cal H}^2} =\frac{p^2}{{\cal
 H}^2}\left[ \vk_1 + \vk_2 \left( \frac{p}{a M_*} \right)^2+
 \vk_3 \left( \frac{p}{a M_*} \right)^4 \right] 
 \,,
\label{LSdisp}
\end{align}
where ${\cal{H}} =a'/a = aH$. The mode equation is given by 
\begin{align}
 & \varphi_p'' +  2{\cal{H}}\varphi_p' + \omega^2_{\varphi}\varphi_p = 0\,.
\end{align}
During inflation, we have
\begin{align}
 & {\cal{H}} =-1/t\,,  \label{Eq:cH}
\end{align}
where we have neglected the corrections suppressed by the slow-roll
parameters.  
When the contribution from either of $z=1,\, 2,\, 3$ dominates the others, 
using Eq.~(\ref{Eq:cH}) and imposing the adiabatic initial condition: 
\begin{align}
 & \varphi_p(t) \to  \frac{1}{a} \frac{1}{\sqrt{2
 \omega_\varphi}} e^{- i \int \dd t\, \omega_\varphi}\,,
\end{align}
we can solve the mode equation as 
\begin{align}
\varphi_p = \frac{1}{2a}\sqrt{\frac{-\pi t}{z}}
e^{i\frac{\pi(2\nu+1)}{4}}H^{(1)}_\nu
 \left[\frac{\sqrt{\vk_z}}{z} \frac{p}{{\cal H}} \left(
 \frac{p}{a M_*} \right)^{z-1} \right]\,,\label{phip}
\end{align}
where the index of the Hankel function is given by
\begin{align}
\label{nu}
\nu= \frac{3}{2z} \;.
\end{align}
At the Hubble crossing, $\omega_\varphi/{\cal H} \simeq z$, we obtain
the power spectrum 
of Lifshitz scalar $\varphi$ as
\begin{align}
 & {\cal P}_{LS}(p)  \equiv \frac{p^3}{2\pi^2} \left| \varphi_p \right|^2 =
 \frac{\a_1^{\nu(z-1)}}{\vk_z^\nu}\frac{(2^\nu \Gamma[\nu])^2}{8\pi^3} z^{\frac{3}{z}-1}
  M_P^2\left( \frac{H_{p}}{M_P} \right)^{\frac{3}{z} - 1} \,,  \label{Exp:Pvarphi}
\end{align}
where $H_p$ denotes the Hubble parameter at this time. In order to
obtain the power spectrum at the end of inflation, we need to solve the
time evolution also after $\omega_\varphi/{\cal H} \simeq z$.  
In the massless case  
$\varphi$ stops evolving in time soon after the Hubble crossing.  
Then Eq.~(\ref{Exp:Pvarphi}) gives the spectrum of $\varphi$ at the end
of inflation. Notice that, as discussed in
Ref.~\cite{Mukohyama:2009gg}, 
for $z=3$ the spectrum of Lifshitz scalar is exactly flat. This is a
consequence of the fact that for $z=3$ the scaling dimension of the
scalar $\varphi$ vanishes.
If the Lifshitz scalar has a small mass, its evolution must also be
traced after the Hubble crossing and the final spectrum in general
depends on the details of this evolution.

\subsection{Gravitational waves}
\label{ssec:gw}
In this subsection we compute the spectrum of the gravitational waves
generated during inflation in HL gravity. We consider the metric, 
\begin{align}
 & N= 1, \quad N_i = 0, \quad  h_{ij} = a^2 \left( \delta_{ij} + \gamma_{ij} \right)
\end{align}
with the transverse traceless condition on the perturbations:
\begin{align}
 & \partial_i \gamma_{ij} =0 \,, \qquad \gamma_{ii}=0\,. 
\end{align}
The quadratic Lagrangian density for the gravitational waves is given by
\begin{align}
 &  {\cal L}_{GW} = \frac{M_*^2}{8 \alpha_1}a^2 \biggl[
 {\gamma '}^i\!_j {\gamma '}^j\!_i - \frac{\alpha_1}{\alpha_3}
  \partial_k \gamma^i\!_j \partial^k \gamma^j\!_i 
-\frac{\alpha_1}{\beta_1 M_*^2} a^{-2} \partial^2 \gamma^i\!_j
 \partial^2 \gamma^j\!_i -  \frac{\alpha_1}{\gamma_1 M_*^4}a^{-4}
 \partial^2 \partial_k \gamma^i\!_j \partial^2 \partial^k \gamma^j\!_i
 \biggr] . \label{actionGW}
\end{align}
This is the most general form of the Lagrangian for 
linear tensor perturbations in HL gravity in the absence
of parity violation and non-minimal coupling to the inflaton.
(See Ref.~\cite{TakahashiSoda} for the 
computation of the polarized gravitational wave spectrum in the presence
of the parity violation.)

Taking variation with respect to $ \gamma_{ij}$, we obtain the mode
equation for $\gamma_{ij}$ as usual,
\begin{align}
 & \gamma_{ij\, p}'' + 2{\cal{H}}  \gamma_{ij\, p}' +
 \omega^2_\gamma \,  \gamma_{ij\, p}=0\,,
\end{align} 
where the frequency $\omega_{\gamma}$ is given by
\begin{align}
 & \frac{\omega^2_{\gamma}(\eta, p)}{{\cal{H}}^2} = \left(
   \frac{p}{{\cal{H}}} \right)^{\!2}\,
 \sum_{z=1}^3 \vk_{\gamma,z} \left( \frac{p}{a M_*} \right)^{2(z-1)} 
\, ,
\end{align}
with $\vk_{\gamma,z}$ given in Eq.~(\ref{Def:vk}). We quantize the
gravitational waves as 
\begin{align}
 &  \gamma_{ij}(x) = \sum_{\lambda=\pm} \int \frac{\dd^3
 \bm{p}}{(2\pi)^{3/2}} \gamma_p (t)
 e^{(\lambda)i}\!_j(\bm{p}) e^{i \sbm{p} \cdot \sbm{x}}
 a_{\sbm{p}}^{(\lambda)} + ({\rm h.c.})\,, \label{Exp:gI}
\end{align}
where $\lambda$ is the helicity of the gravitational waves,
$e^{(\lambda)}_{ij}$ are the standard transverse and traceless 
polarization tensors, and $a_{\sbm{k}}^{(\lambda)}$ are the annihilation operators which satisfy
\begin{eqnarray}
  \left[ a_{\sbm{k}}^{(\lambda)},\, a_{\sbm{p}}^{(\lambda')\dagger}
  \right] = \delta_{\lambda \lambda'} \delta^{(3)} (\bm{k} - \bm{p})\,. 
\end{eqnarray}
The number of the polarizations in HL gravity is the same as in GR.
Imposing the adiabatic initial condition:
\begin{align}
 & \gamma_p(t) \to \frac{2}{a M_P} \frac{1}{\sqrt{2
 \omega_\gamma}} e^{- i \int \dd t\, \omega_\gamma}\,,
\end{align}
we obtain the mode functions $\gamma_p$ as
\begin{align}
 & \gamma_p(t) =  \frac{1}{M_P a}\sqrt{\frac{-\pi t}{z}} 
e^{i \frac{\pi (2\nu +1)}{4}} H_{\nu}^{(1)}  \left[
\frac{\sqrt{\vk_{\gamma,z}}}{z} \frac{p}{{\cal H}} \left(
 \frac{p}{a M_*} \right)^{z-1} \right]\,,
\label{gammap}
\end{align}
where the Hankel index $\nu$ is given in Eq.~(\ref{nu}).
Like in the GR, $\gamma_p$ is conserved in time for
$\omega_\gamma/{\cal H} < z$. Using Eq.~(\ref{gammap}) we obtain the power spectrum of the gravitational waves as
\begin{align}
 & {\cal P}_\gamma \equiv \frac{p^3}{\pi^2} \left| \gamma_p \right|^2 =
 \frac{\alpha_1^{\nu (z-1)}}{\vk_{\gamma,z}^{\nu}}
 \frac{(2^{\nu} \Gamma[\nu])^2}{\pi^3} z^{\frac{3}{z}-1}
 \left( \frac{H_{p, \gamma}}{M_P} \right)^{\frac{3}{z} - 1} \,,\label{Exp:powertens}
\end{align}
where $H_{p, \gamma}$ denotes the Hubble parameter when
$\omega_\gamma/{\cal H} \simeq z$. 

The spectral index for the gravitational waves is given by 
\begin{align}
 & n_t  \equiv \frac{\dd \ln {\cal{P}}_{\gamma}}{\dd \ln p} \simeq -
  \frac{3-z}{z} \varepsilon_1 \,. \label{Exp:nt}
\end{align}
In 4D Diff invariant theory, the spectrum of the primordial gravitational waves
is generically red-tilted in an inflationary universe with
$\varepsilon_1 > 0$~\cite{CGNV}. By contrast, in HL gravity, for
$z=3$, the spectral index $n_t$ vanishes even if $\varepsilon_1 \neq 0$. 
This serves as a distinctive feature of the anisotropic scaling
regime of gravity.  
Since the lapse function is irrelevant to the
gravitational waves at the linear order of perturbation, the results
of this section apply both to the projectable
and non-projectable versions of HL gravity.

\section{Decoupling and non-decoupling of khronon}
\label{Sec:Khronon}
In this section, we consider the scalar linear perturbations including
the inflaton and metric perturbations. We express the fields as,
\begin{align}
& \Phi(t,\, \bm{x}) = \phi(t) + \varphi(t,\, \bm{x})\,,
\quad N=a(1+\delta N)\,,\quad N_i=a^2\partial_iB\;,
 \quad h_{ij} = a^2e^{2\cR} \delta_{ij}\,. \label{Exp:ADMmetric}
\end{align}
In general relativity, the
metric perturbation $\cR$ and the fluctuation of
the inflaton $\varphi$ are not independent. By contrast, 
in HL gravity ${\cal R}$ serves an
additional scalar degree of freedom, khronon, as a consequence of the
lack of 4D Diff invariance. In this section we discuss the evolution of
khronon both in the projectable
and non-projectable versions of HL gravity. We will find that the khronon
behaviour differs qualitatively in these two cases.

\subsection{Projectable HL gravity}
\label{ssec:P}
First we consider the projectable version of HL gravity. A review of this
version can be found in
Ref.~\cite{Mukohyama:2010xz}. 
In this case the lapse function is constrained to be homogeneous and
does not affect local physics. Setting $\delta N=0$ 
and integrating out the non-dynamical field $B$ we find the action, 
\begin{align}
\label{genLagRf}
S = \int \dd t \int \dd^3 \bm{p} 
\left[ {\cal{L}}_{\cR} + {\cal{L}}_{\varphi} + {\cal{L}}_{\cR\varphi}\right]\,, 
\end{align}
with
\begin{align}
  &{\cal L}_{\cR} =  a^2M_*^2 \frac{1+
 \bar{\alpha}}{\alpha_1\bar{\alpha}} \left[{\cR '}_{\sbm{p}}
 {\cR '}_{- \sbm{p}} - \omega_{\cR}^2(t,\,p)  \cR_{\sbm{p}}
 \cR_{- \sbm{p}} \right]\,,  \\
  & {\cal L}_{\varphi} = \frac{a^2}{2} \left[{\varphi '}_{\sbm{p}}
 {\varphi '}_{- \sbm{p}} - \omega_{\varphi}^2(t,\,p)  \varphi_{\sbm{p}}
 \varphi_{- \sbm{p}} \right]\,, \\
  &  {\cal L}_{\cR\varphi} =a^2
 \frac{1-2\bar{\alpha}}{\bar{\alpha}} {\phi '}\varphi_{\sbm{p}}{\cR '}_{-\sbm{p}} \,.  \label{mix_pro}
\end{align}
The frequencies $\omega^2_\cR$ and $\omega^2_\varphi$ are given by
\begin{align}
\label{omegaR_pro}
 & \frac{\omega^2_{\cR}(t,\,p)}{{\cal{H}}^2} =  \frac{\alpha_1\bar{\alpha}
 }{1+ \bar{\alpha}} \biggl(\frac{p}{{\cal{H}}}\biggr)^2  \left[ -
 \frac{1}{\alpha_3} + \biggl( \frac{3}{\beta_1} + \frac{8}{\beta_2}
 \biggr)  \biggl( \frac{p}{a M_*} \biggr)^2+ \biggl( \frac{3}{\gamma_1} + \frac{8}{\gamma_2}
 \biggr)  \biggl( \frac{p}{a M_*} \biggr)^4 \right]\,,\\ 
 &\frac{\omega^2_{\varphi}(t,\,p)}{{\cal{H}}^2} =  
 \biggl(\frac{p}{{\cal{H}}}\biggr)^2 \left[ \vk_1 + \vk_2 \biggl(
 \frac{p}{a M_*} \biggr)^2 + \vk_3 \biggl( \frac{p}{a M_*} \biggr)^4  \right]
  -\frac{1+\bar{\alpha}}{\bar{\alpha}}\varepsilon_1+ \frac{3(1+\bar{\alpha})}{1-2\bar{\alpha}}\eta_V\,.  
\end{align}
We observe that in a de Sitter universe, where the inflaton is absent,
khronon $\cR$ behaves as a massless Lifshitz scalar and thus is conserved at
super Hubble scales. However, mixing with the inflaton (\ref{mix_pro})
essentially modifies the dynamics.

Positivity of khronon kinetic energy requires,
\begin{align}
 &  \frac{1+ \bar{\alpha}}{\alpha_1\bar{\alpha}} > 0\,,
\end{align}
which implies that $\cR$ suffers from a gradient instability in the IR
limit, since $\alpha_3 > 0$. 
An attempt to suppress this instability by taking the coefficient 
$\alpha_1 \bar{\alpha}/\alpha_3(1+ \bar{\alpha})$ to be small leads to
strong coupling and invalidates the perturbative description 
(see a detailed discussion in Ref.~\cite{Blas:2010hb}).
Thus, projectable HL gravity cannot provide a viable low-energy
phenomenology in the regime of weak coupling. By analogy with
non-Abelian gauge theories, one might envision a scenario where strong
coupling occurs only in IR and leads to confinement of khronon at low
energies. However, currently there exist no controllable
realizations of this scenario. Here we restrict to the anisotropic
scaling regime where the second and third terms in the brackets in
(\ref{omegaR_pro}) dominate, the theory is stable and weakly coupled.  

Estimates of various terms in the Lagrangian show that at 
$\omega_{\cR},\omega_{\varphi}\gg {\cal H}\sqrt{\varepsilon}$
the mixing term between $\cR$ and $\varphi$ is negligible and these two
fields evolve independently. 
Assuming that either $z=2$ or $z=3$ contribution is
dominant and imposing the standard WKB initial condition we find the
mode functions for $\cR$ and $\varphi$,  
\begin{align}
&\cR_p(t) = \frac{1}{M_P a}\,\sqrt{\frac{\bar{\alpha}}{1+ \bar{\alpha}}}\,
  \sqrt{\frac{-\pi t}{8z}}
 e^{i\frac{\pi(2\nu+1)}{4}}
H^{(1)}_{\nu}\left[ \frac{\omega_{\cR}}{z{\cal H}} \right]\,, 
 \label{Khronon_pro}\\    
&\varphi_p(t) =\frac{1}{2a}\,
 \sqrt{\frac{-\pi t}{z}}e^{i\frac{\pi(2\nu+1)}{4}}
H^{(1)}_{\nu}\left[\frac{\omega_{\varphi}}{z{\cal H}}
 \right]\,,  
\end{align}
where the Hankel index $\nu$ is given by Eq.~(\ref{nu}). 
We did not write explicitly the arguments of the Hankel functions;
they have the same dependence on $p$ and $t$ as those 
in Eqs.~(\ref{phip}) and (\ref{gammap}). We have also 
neglected the slow-roll correction in 
$\omega_\varphi/{\cal H}$, since the momentum dependent contribution
dominates it in this regime. 

The above solutions cannot be extended to super Hubble evolution where 
$\omega_{\cR},\omega_{\varphi}\lesssim {\cal H}\sqrt{\varepsilon}$
and the mixing between $\cR$ and $\varphi$ becomes important.
In this regime we have two light Lifshitz scalars, the inflaton and
khronon, which are mixed with each other. As in a 4D Diff invariant
theory with more than one light scalar fields, in this case we do not
find an adiabatic mode which is conserved in time at large
scales (see, e.g., Ref.~\cite{Gordon:2000hv}). Then, in order to
compute the observed fluctuations, we need to 
solve the time evolution which can depend on concrete models of the
reheating, the transition to the isotropic scaling regime, and so on. 
As discussed above, this would require also a controllable description
of the mechanism that suppresses the IR instability of the theory, which is
currently missing. 
Therefore it appears problematic to provide a robust prediction for
primordial scalar power spectrum
in the projectable version of HL gravity.

\subsection{Non-projectable HL gravity}
\label{ssec:NP}
In this subsection we will find that the time evolution of khronon in
the non-projectable version is qualitatively different from the one in
the projectable version discussed above. In
particular, we will show that khronon is decoupled from the adiabatic
curvature perturbation $\zeta$ at large 
scales. Because of that, $\zeta$ is conserved in time as in the single field
model with 4D Diff invariance. Therefore, we can derive a
robust prediction for the power spectrum of $\zeta$ without solving
the detailed evolution after the Hubble crossing.

\subsubsection{Mass gap of khronon and anti-friction}
\label{sssec:mgap}
In the non-projectable version of HL gravity, upon eliminating the
non-dynamical fields $B$ and $\delta N$, we obtain the action for  
$\cR$ and $\varphi$ in the form (\ref{genLagRf})
with
\begin{align}
 & {\cal L}_{\cR} = a^2  M_*^2 \frac{1+
 \bar{\alpha}}{\alpha_1\bar\a} \left[(1- \Omega_1(t,\, p))  \cR'_{\sbm{p}}
 \cR'_{- \sbm{p}} - \omega_{\cR}^2(t,\,p)  \cR_{\sbm{p}}
 \cR_{- \sbm{p}} \right]\, \label{mixLagMB}\;.
\end{align}
The expressions for 
${\cal L}_\varphi, {\cal L}_{\cR\varphi}$ are given in
Appendix \ref{app:A}. We introduce the functions 
$\Omega_i(t,\, p)$ with $i=1,\,2$ as
\begin{align}
 & \Omega_1(t,\, p) = \left\{  1 + \frac{\bar{\alpha}\varepsilon_1}{1- 2
 \bar{\alpha}} +  \frac{\alpha_1 \bar{\alpha}}{2(1+ \bar{\alpha})}
 \biggl(\frac{p}{{\cal{H}}} \biggr)^2
 \biggl[1+ \frac{1}{\beta_4} \biggl( \frac{p}{a M_*} \biggr)^2  +
 \frac{1}{\gamma_4}  \biggl( \frac{p}{a M_*} \biggr)^4 \biggr]
 \right\}^{-1}, \label{Def:Omega1} \\ 
 & \Omega_2(t,\,p) = \frac{\alpha_1 \bar{\alpha}}{1+ \bar{\alpha}}
  \biggl(\frac{p}{{\cal{H}}} \biggr)^2 \left[ - \frac{1}{\alpha_3} + \frac{1}{\beta_3}  \biggl(
 \frac{p}{a M_*} \biggr)^2 + \frac{1}{\gamma_3}  \biggl(
 \frac{p}{a M_*} \biggr)^4 \right] \Omega_1(t,\, p)\,. 
\label{Def:Omega2}
\end{align}
In terms of these quantities the frequency $\omega_{\cR}$ is expressed as,
\begin{align}
  \frac{\omega^2_{\cR}(t,\,p)}{{\cal{H}}^2} =&  \frac{\alpha_1\bar\a
 }{1+ \bar{\alpha}} \biggl(\frac{p}{{\cal{H}}}\biggr)^2  \left[ -
 \frac{1}{\alpha_3} + \biggl( \frac{3}{\beta_1} + \frac{8}{\beta_2}
 \biggr)  \biggl( \frac{p}{a M_*} \biggr)^2+ \biggl( \frac{3}{\gamma_1} + \frac{8}{\gamma_2}
 \biggr)  \biggl( \frac{p}{a M_*} \biggr)^4 \right] \cr
  &  + 
 \frac{\Omega_2^2(t,\,p)}{\Omega_1(t,\, p)}
 -  \frac{( a^2 {\cal H} \Omega_2(t,\,p))'}{ a^2
 {\cal H}^2} . \label{Exp:omegaRMB} 
\end{align}
We observe that the khronon Lagrangian is now much more complicated
than in the projectable case. A crucial new feature is the dependence
of the coefficient in front of the term with time derivatives in
(\ref{mixLagMB}) on the mode momentum. This leads to a peculiar
behavior of khronon in inflationary universe, as we presently discuss. 

It is convenient to introduce the notation,
\begin{align}
  X(t,p) &\equiv  \biggl(\frac{p}{{\cal{H}}} \biggr)^2
 \biggl\{1+ \frac{1}{\beta_4} \biggl( \frac{p}{a M_*} \biggr)^2  +
 \frac{1}{\gamma_4}  \biggl( \frac{p}{a M_*} \biggr)^4 \biggr\} \,.
\label{Xdef}
\end{align}
Roughly speaking, this quantity characterizes the (square of the)
ratio between the frequencies of the perturbations and the Hubble rate
(the precise expressions will be given below). In the course of
cosmological evolution it goes through four different regimes:
\begin{align}
\label{Xreg1}
&\text{{\it (a) }} \qquad X\gg 1/\a^2\;,\\
\label{Xreg2}
&\text{{\it (b) }} \qquad \varepsilon_1/\a\ll X\ll 1/\a^2\;,\\
\label{Xreg3}
&\text{{\it (c) }} \qquad 1\ll X\lesssim\varepsilon_1/\a\;,\\
\label{Xreg4}
&\text{{\it (d) }} \qquad X\ll 1\;.
\end{align}
Recall that in the non-projectable case the parameters
are assumed to satisfy the hierarchy (\ref{parameterNP}). Here and
below we symbolically denote the small quantities in
(\ref{parameterNP}) by $\a$. Additionally, we will assume for the
moment that 
\begin{equation}
\label{epsalhier}
\ve_1/\a\gg 1\;.
\end{equation}
The opposite case will be commented on at the end of the section. Let us
consider the above regimes one by one.

{\it (a)} $X\gg 1/\a^2$. In this case
\begin{gather}
\Omega_1 \simeq   \frac{1}{\a_1\bar\a X}\ll 1\;,\\
\frac{\omega_\cR^2}{{\cal H}^2}\simeq 2\a_1\bar\a
\bigg(\frac{p}{\cal H}\bigg)^2\frac{\left[  -
 \frac{1}{\alpha_3} + \frac{1}{\beta_3}  (
 \frac{p}{aM_*})^2 + \frac{1}{\gamma_3}  (
 \frac{p}{aM_*})^4 \right]^2 }{1+ \frac{1}{\beta_4}  (
 \frac{p}{aM_*})^2 + \frac{1}{\gamma_4}  (
 \frac{p}{aM_*} )^4 }\;.\label{omegaRUV}
\end{gather}
In the latter expression we recognize the dispersion relation of
khronon in flat spacetime (\ref{Exp:DRNP}) (up to suppressed
corrections). 
In the UV regime,
$p>aM_*$, it behaves as a Lifshitz scalar with $z=3$ and
\[
\omega_\cR^2 \simeq \frac{2\g_4\a_1\bar\a}{\g_3^2}p^2\Big(\frac{p}{aM_*}\Big)^4\;,
\]
whereas if $p<aM_*$ (but (\ref{Xreg1}) still satisfied) it obeys
the $z=1$ scaling,
\[
\omega_\cR^2\simeq \frac{2\a_1\bar\a}{\a_3^2}p^2\;.
\]
Note that in both limiting cases the ratio $\omega_\cR^2/{\cal H}^2$
is of order $X(t,p)$. From expressions given in Appendix~\ref{app:A}
one can infer that inflaton $\varphi$ also behaves in this regime as a
Lifshitz scalar with dispersion relation (\ref{LSdisp}). Further, by
estimating various terms in the Lagrangian ${\cal L}_{\cR\varphi}$,
Eq.~(\ref{mixLag}), it is straightforward to check that mixing between
modes $\cR$ and $\varphi$ is negligible\footnote{Strictly speaking,
  the mixing can be resonantly enhanced if the frequencies
  $\omega_\cR(t,p)$ and $\omega_\varphi(t,p)$ happen to cross at some
  specific time. In our analysis, we do not consider this
  possibility. However, even if the crossing takes place, the mode
  functions stay in the WKB form and the time evolution remains
  essentially unchanged after the crossing.}.  

The IR limit $p\ll aM_*$ of non-projectable HL gravity is closely
related to Einstein-aether theory~\cite{Jacobson:2010mx}.
Evolution of cosmological perturbations in the latter theory was
analyzed in~\cite{APSG} and 
it was shown that in the short wavelength limit the khronon $\cR$ and the
fluctuation of the inflaton $\varphi$ are decoupled from each 
other, which allows to impose the WKB initial condition as usual. 
Our analysis provides a generalization of this result to the UV modes
of HL gravity where terms with Lifshitz scaling $z=2$ and $3$ are
important.  

{\it (b)} $\ve_1/\a\ll X\ll 1/\a^2$. In this regime we have,
\[
\Omega_1\approx 1-\frac{\a_1\bar\a}{2}X\;,
\]
and the khronon Lagrangian takes the form,
\[
{\cal L}_\cR=a^2\frac{M^2_*X(t,p)}{2}\big(\cR'_{\sbm p}\cR'_{-{\sbm p}}
-\bar\omega_\cR^2(t,p)\cR_{\sbm p}\cR_{-{\sbm p}}\big)\;.
\]
The khronon frequency $\bar\omega_\cR$ now reads,
\begin{equation}
\frac{\bar\omega_{\cR}^2}{{\cal H}^2}=\!
\frac{a^2\Big[m_{k,1}^2\!
+\!\frac{m_{k,2}^2}{\b_4}(\frac{p}{aM_*})^2
\!+\!\frac{m_{k,3}^2}{\g_4}(\frac{p}{aM_*})^4\Big]}{
{\cal H}^2\left[1+\frac{1}{\b_4}(\frac{p}{aM_*})^2
+\frac{1}{\g_4}(\frac{p}{aM_*})^4\right]}
+2\a_1\bar\a
\bigg(\frac{p}{\cal H}\bigg)^{\!\!2}\frac{\left[  -
 \frac{1}{\alpha_3} + \frac{1}{\beta_3}  (
 \frac{p}{aM_*})^2 + \frac{1}{\gamma_3}  (
 \frac{p}{aM_*})^4 \right]^2 }{1+ \frac{1}{\beta_4}  (
 \frac{p}{aM_*})^2 + \frac{1}{\gamma_4}  (
 \frac{p}{aM_*} )^4 },
\label{omegaRbar}
\end{equation}
where
\begin{align}
\label{khmass1}
 & m_{k,1}^2 \equiv 2  \frac{\varepsilon_1}{\alpha_3} H^2 \,, \\
\label{khmass2}
 & m_{k,2}^2 \equiv 2  \beta_4 \left( \frac{3}{\beta_1} +
 \frac{8}{\beta_2} + \frac{1}{\beta_3} \right) H^2 \,, \\
\label{khmass3}
 &  m_{k,3}^2 \equiv 2 \gamma_4 \left( \frac{3}{\gamma_1} +
 \frac{8}{\gamma_2} + \frac{3}{\gamma_3} \right) H^2\,.
\end{align}
The second term in (\ref{omegaRbar}) is the same as
(\ref{omegaRUV}). However, we observe that a new contribution appears
which gives the khronon a mass gap. For the regime where terms with a
given $z=1,2,3$ dominate the mass is given by $m_{k,z}$. Notice that
$m_{k,1}$ is suppressed
by the slow-roll parameter
compared to $m_{k,2}$ and $m_{k,3}$. Still, within our assumption
(\ref{epsalhier}) all the masses are parametrically larger than the
Hubble rate.

Inspection of the terms ${\cal L}_{\varphi}$ and ${\cal L}_{\cR\varphi}$
in the Lagrangian (see Appendix~\ref{app:A}) again shows
that the inflaton fluctuation $\varphi$ behaves as a Lifshitz scalar
with dispersion relation (\ref{LSdisp}) and decouples from
$\cR$. Thus we can study evolution of $\cR$ separately.

Due to the mass gap, the khronon rapidly oscillates. However, unlike
one could naively expect, the
amplitude of these oscillations does not decay in the Lifshitz
regime. When either 
$z=2$ or $z=3$ contribution is dominant and the khronon frequency
$\bar\omega_\cR$ is dominated by the mass term, as happens for 
$X\ll 1/\a$,
the equation for $\cR$ reads,
\begin{align}
 & \cR'' - 2 (z-1) {\cal H} \cR' + a^2 m^2_{k,\,z} \cR =0\,.  \label{Eq:KhronondS}
\end{align}
The second term here produces an `anti-friction'. The
canonically normalized mode functions have the form in the WKB approximation,
\begin{align}
 & \cR(t) =\frac{H M_*^{z-2}}{p^z\sqrt{2m_{k,z}}}\,
(a(t))^{z-3/2} \, e^{- i \int \dd t a(t) m_{k,z}}\,, \label{Exp:cRWKB}
\end{align}
and describe oscillations with a growing amplitude, $|\cR_p|\propto
a^{z-3/2}$. We are going to see in the next subsection that the growth
of khronon perturbations persists also at $X<\ve_1/\a$ as long as the
modes remain in the Lifshitz regime and stops only when they
pass into the isotropic scaling $z=1$. To stay within the validity
of perturbation theory, we will impose the requirement that the
amplitude of khronon perturbations remains small throughout the
cosmological evolution, $p^{3/2}|R_p|<1$. This translates into certain
conditions on the inflationary parameters that will be discussed below.   

\subsubsection{Khronon-inflaton mixing}
As the modes are further redshifted, the fields $\cR$ and $\varphi$
get mixed and no longer provide a convenient basis for
perturbations. To find the appropriate basis, we study the Lagrangian
for $\cR$ and $\varphi$ in the regime:

{\it (c)} $1\ll X\lesssim \ve_1/\a$. We will focus in this subsection 
on the case when the terms with Lifshitz
scaling $z=2$ or $3$ dominate in the dispersion relation. This is true
if the inflationary Hubble rate $H$ is bigger than $M_*$, which is
the scenario of primary interest to us. For completeness we consider
the case of isotropic scaling 
 in Appendix~\ref{app:iso}.

In the Lifshitz regime the leading mixing term is the first
contribution in (\ref{mixLag}). Simplifying the expressions using the
assumed parameter hierarchy and introducing the canonically normalized
field 
\[
\hat \cR\equiv \sqrt{2\ve_1+\a_1X}\,M_P \cR\;,
\]  
we obtain the relevant part of the Lagrangian,
\begin{equation}
\label{Lagmixed}
  {\cal L}=\frac{a^2}{2}\big(\hat \cR'_{\sbm{p}}\hat\cR'_{-\sbm{p}} - 
\hat\omega_{\cR}^2 \hat\cR_{\sbm{p}}\hat\cR_{-\sbm{p}}\big)
+\frac{a^2}{2} \big( \varphi'_{\sbm{p}} \varphi'_{- \sbm{p}} -
 \omega^2_{\varphi}  \varphi_{\sbm{p}}  \varphi_{-\sbm{p}}\big)
 -\frac{a^2}{\sqrt{1+\frac{\a_1 X}{2\ve_1}}}
 {\varphi '}_{\sbm{p}} {\hat\cR '}_{- \sbm{p}}\;, 
\end{equation}
where $\omega_\varphi^2$ is given by (\ref{LSdisp}) and 
\[
\hat\omega_\cR^2=\frac{\omega_\cR^2}{\bar\a(\ve_1+\a_1X/2)}\;.
\]
In deriving these expressions we have neglected contributions of order
${\cal H}$ into the frequencies. Note that $\hat\omega_\cR$ is much
higher than $\omega_\varphi$. Indeed, we have
\begin{equation}
\label{omcromvarphi}
\hat\omega_\cR^2\simeq a^2\frac{\a_1}{\ve_1}m_{k,z}^2 X \simeq {\cal
  H}^2\frac{X}{\ve_1} \gg {\cal H}^2 X \simeq \omega_\varphi^2\;. 
\end{equation}

\begin{figure}[ht]
 \centering
  \includegraphics[width=12.0cm, bb=0 0 750 850]{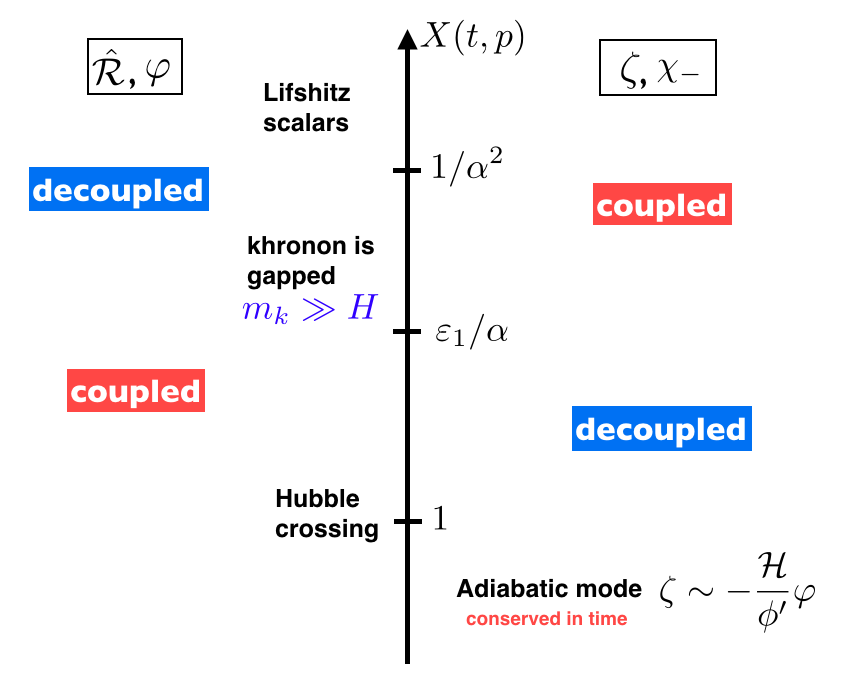}
  \caption{Summary of the time evolution of the fluctuations. The
    central axis denotes the quantity $X(t,p)$ introduced in Eq.~(\ref{Xdef}). }
    \label{fig_HL}
\end{figure}

The Lagrangian (\ref{Lagmixed}) confirms explicitly our previous
assertion that at $X\gg \ve_1/\a$ the mixing between $\cR$ and
$\varphi$ is negligible. On the other hand, we see that at $X\ll
\ve_1/\a$ it becomes essential. To identify the independent modes we use
the substitution,
\begin{align}
&\chi_+ =   \varphi\cos{\theta} -\hat\cR \frac{\hat\omega_\cR}{\omega_\varphi}
\sin{\theta}  \,,  \label{Def:chi+} \\
&\chi_- = \varphi\frac{\omega_\varphi}{\hat\omega_\cR}\sin{\theta} + 
\hat\cR \cos{\theta}\, , \label{Def:chi-}
\end{align}
and find that the mixing term between $\chi_{\pm}$ disappears provided that\footnote{We
  again neglect contributions proportional to ${\cal H}$ that come from
time variation of $\hat\omega_\cR$, $\omega_\varphi$, $X$. These are
irrelevant as long as the frequencies of the fields are higher than the
Hubble rate.} 
\begin{align}
\tan{2\theta}
=\frac{2\omega_\varphi\hat\omega_\cR}{\hat\omega_\cR^2-\omega_\varphi^2}
\frac{1}{\sqrt{1 + \frac{\alpha_1X}{2\ve_1}}}\,. \label{theta}  
\end{align}
Due to (\ref{omcromvarphi}) 
the mixing angle $\theta$ is always small and the expressions for
the new variables $\chi_\pm$ simplify.
At $X\ll \ve_1/\a$ they become,
\begin{align}
\label{chi+simp}
&\chi_+=\varphi-\sqrt{2\ve_1}M_P\cR\;,\\
\label{chi-simp}
&\chi_-=\sqrt{2\ve_1}M_P\cR
+\Big(\frac{\omega_\varphi}{\hat\omega_\cR}\Big)^2\varphi\;,
\end{align} 
where we have switched back to the original metric perturbation
$\cR$. In the expression (\ref{chi+simp}) we recognize the standard
gauge invariant variable
\begin{align}
 & \zeta \equiv {\cal R} - \frac{\cal H}{\phi'}\,\varphi=
-\frac{\cal H}{\phi'}\chi_+ 
 \label{Def:zeta}
\end{align}
describing curvature perturbation on the slices of constant inflaton
field. The Lagrangian for $\chi_\pm$ reads,
\begin{align}
 & {\cal L} =  \frac{a^2}{2} 
 \big( {\chi_{+}'}^2 - \omega_\varphi^2(t,p) \chi_{+}^2\big)
+a^2\,\frac{\a_1X(t,p)}{4\ve_1} \big( {\chi_{-}'}^2 
- \bar\omega_\cR^2(t,p) \chi_{-}^2\big)\;,
 \label{diagLag} 
\end{align}
where $\bar\omega_\cR^2$ is the same as in (\ref{omegaRbar}). We see that
$\chi_+$ (or equivalently $\zeta$) inherits the dispersion relation of
the inflaton, whereas the second mode $\chi_-$ --- that of khronon. In
other words, in the regime (\ref{Xreg3}) we still have two independent
physical excitations, inflaton and khronon, with their respective
dispersion relations (\ref{LSdisp}), (\ref{omegaRbar}). The
corresponding eigenfunctions are connected to the original variables
by (\ref{chi+simp}), (\ref{chi-simp}). This is illustrated in
Fig.~\ref{fig_HL}. 

\subsubsection{Long wavelength evolution and power spectrum}
We have shown that the variables $\chi_\pm$ are independent as long as
the frequencies of the modes remain higher than the Hubble rate. As
the modes redshift and approach the `horizon crossing', $X\simeq 1$, the
situation gets more complicated due to the terms proportional to
${\cal H}$ in the Lagrangian that can no longer be neglected. However,
the situation simplifies again for `super Hubble' modes corresponding
to the regime:

{\it (d)} $X\ll 1$. In the standard relativistic single field
inflation the curvature perturbation $\zeta$ is conserved at these
scales. In Appendix~\ref{app:A} we show that this also holds for
non-projectable HL gravity, despite the presence of khronon, by
explicitly writing the quadratic Lagrangian in terms of $\zeta$ and
$\varphi$. All non-derivative terms in the $\zeta$-equation turn out
to be suppressed by $X$, so that we obtain the solution,
\begin{equation}
\label{zetasuper}
\zeta=const\;.
\end{equation}   
This allows to immediately write down the power spectrum for $\zeta$
by matching to the amplitude of $\chi_+$ fluctuations at the Hubble
crossing, see Eq.~(\ref{Def:zeta}),
\begin{align}
 & {\cal P}_\zeta (p) = \frac{1}{2 \varepsilon_{1, p} M_P^2}\, {\cal
 P}_{LS}(p)\,, \label{Exp:PSzetaH}
\end{align}
where ${\cal P}_{LS}(p)$ is the power spectrum of the Lifshitz scalar
and $\varepsilon_{1,p}$ is the value of the slow-roll parameter 
at the Hubble crossing time of the mode $p$. Explicitly we have,
\begin{align}
 & {\cal P}_\zeta (p) = 
\frac{\alpha_1^{\nu (z-1)}}{\varepsilon_{1, p}\vk_z^\nu}
 \frac{(2^\nu \Gamma[\nu])^2}{16 \pi^3} z^{\frac{3}{z}-1}
 \left( \frac{H_{p}}{M_P} \right)^{\frac{3}{z} - 1}\,, 
\qquad\qquad
\nu=\frac{3}{2z}\;.
 \label{Exp:powerzeta0}
\end{align}
Note that for $z=3$ the spectrum is independent of the Hubble rate at inflation,
\begin{equation}
\label{Pzeta3}
{\cal P}_\zeta(p)=\frac{1}{8\pi^2}\frac{\a_1}{\ve_{1,p}\sqrt{\vk_3}}\;,
\qquad\qquad z=3\;.
\end{equation}
The spectral index is given by
\begin{align}
 & n_s -1  \equiv \frac{\dd \ln {\cal{P}}_{\zeta}}{\dd \ln p} = -
 \frac{3-z}{z} \varepsilon_1 -\varepsilon_2 \,,
\end{align}
or alternatively,
\begin{align}
 & n_s - 1 = - \frac{3(1+z)}{z} \varepsilon_V + 2 \eta_V\,. \label{Exp:nsm1}
\end{align}
For $z=1$ we recover the standard expressions.\\

We now analyze the super Hubble behavior of khronon, or `isocurvature'
mode. Despite the fact that the frequency term for $\varphi$ in the
Lagrangian (\ref{tL}), as well as its mixing with $\zeta$, are
suppressed by $X$, it still evolves non-trivially, because its time
derivative term is also proportional to $X$. When the contributions
with Lifshitz scaling $z=2,3$ dominate, the equation for $\varphi$
following from (\ref{tL}), (\ref{Lagzeta2}) simplifies, 
\begin{equation}
\varphi''-2{\cal H}(z-1)\varphi'+a^2m_{k,z}^2
\big(\varphi+\sqrt{2\ve_1}M_P\zeta)=0\;.
\label{varphisup1}
\end{equation}
The combination in brackets in the last term is nothing but 
$\sqrt{2\ve_1}M_P\cR$, which also coincides with $\chi_-$, up to
slow-roll suppressed corrections. Also Eq.~(\ref{varphisup1}) is the
same as the khronon equation (\ref{Eq:KhronondS}). We conclude that
khronon preserves its identity through Hubble crossing. Despite very
long wavelength of the modes, they continue to rapidly oscillate with
growing amplitude
due to anti-friction. The decoupling of $\zeta$ and $\cR$ now
receives an intuitive explanation: these excitations have very
different frequencies and therefore cannot mix.

\begin{figure}[t]
\vspace{-6cm}
\centering
  \includegraphics[width=10.0cm, bb=0 10 700 750]{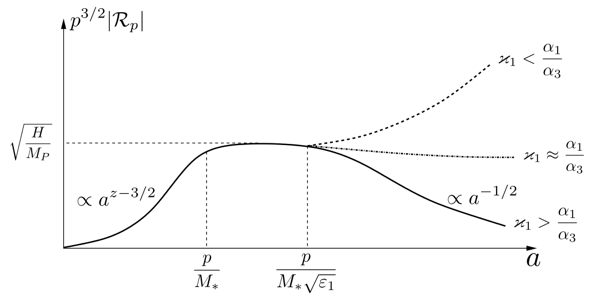} 
  \caption{The amplitude of a khronon mode with conformal momentum $p$
as a function of the scale
    factor. It grows in the Lifshitz regime and reaches the value
    $\sqrt{H/M_P}$. Then it remains constant till 
$a\sim p/M_*\sqrt{\ve_1}$, where the mode enters into the $z=1$
scaling. The subsequent evolution depends on the relation between
velocities of the inflaton and graviton characterized by the
parameters $\vk_1$ and $\a_1/\a_3$, as explained in the main
text. }
    \label{Fig:khronon}
\end{figure}

The amplitude of khronon oscillations seizes to grow when the
momentum redshifts down to $p/aM_*\simeq 1$. For 
$\sqrt{\ve_1}\ll p/aM_*\ll 1$ the equation for $\varphi$ reads,
\begin{equation}
\varphi''+\b_4\frac{p^2m_{k,2}^2}{M_*^2}
\big(\varphi+\sqrt{2\ve_1}M_P\zeta)=0\;,
\label{varphisup2}
\end{equation}
and describes pure oscillations of $\cR$ with constant amplitude. This
is illustrated in Fig.~\ref{Fig:khronon}. Finally, for
$p/aM_*\ll\sqrt{\ve_1}$ the $\varphi$-equation becomes, 
\begin{equation}
\varphi''+\frac{2{\cal H}^2\ve_1}{\a_1}
\Big(\vk_1-\frac{\a_1}{\a_3}\Big)\varphi=0\;.
\label{varphisup3}
\end{equation}
First, we notice that $\varphi$ has completely decoupled from
$\zeta$. This is consistent with the result of \cite{APSG} which
studied inflation in the $z=1$ limit of HL gravity and identified the
independent modes in the super Hubble regime as $\zeta$ and 
$\delta {\cal N}=({\cal H}/\phi') \varphi$. The latter has geometric
interpretation of the difference in the number of $e$-foldings between
the surfaces of constant inflaton (i.e. constant density) and constant
khronon. Second, the nature of solutions to (\ref{varphisup3}) depends
on the sign of the combination in brackets which has the physical
meaning of the difference between (the squares of) the low-energy
velocities of the inflaton, $c_\varphi^2\equiv \vk_1$, and graviton
$c_\g^2\equiv\a_1/\a_3$ (see Eqs.~(\ref{Def:vk}), (\ref{LSdisp})). If
it is positive, the mode $\varphi$ performs rapid oscillations with
the physical frequency $\omega_\varphi/ a \simeq H\sqrt{\ve_1/\a_1}\gg H$
and the amplitude decaying as $a^{-1/2}$. On the other hand, if
$\vk_1<\a_1/\a_3$, the solutions to (\ref{varphisup3}) exhibit an
exponential runaway behavior, signaling an instability. These two
cases are illustrated in Fig.~\ref{Fig:khronon}. To avoid instability,
we will assume that $\vk_1>\a_1/\a_3$. 

Equation (\ref{varphisup3}) has been derived under the assumption that
the low-energy velocities of inflaton and graviton differ by a factor
of order one. Alternatively, one can impose the requirement that this
difference should be small, $c_\varphi^2-c_\g^2={\cal O}(\a)$, which
corresponds to an emergence of approximate Lorentz invariance at low
energies. In this case one must retain additional contributions of the
same order in the expression (\ref{tomegaphi}) for the frequency of
$\varphi$, so that the $\varphi$-Lagrangian becomes,
\begin{equation}
\tilde{\cal L}_\varphi=a^2\frac{\a_1}{4\ve_1}\bigg(\frac{p}{\cal
  H}\bigg)^2
\bigg[{\varphi'}^2-{\cal H}^2\,\ve_1\bigg(1-\frac{\ve_2}{2\ve_1}
+\frac{2(c^2_\varphi-c^2_\g)}{\a_1}
+\frac{3c_{k}^2}{c_\g^2}\bigg)\,\varphi^2\bigg]\;,
\label{varphisup4}
\end{equation}
where $c_k^2\equiv 2\a_1\bar\a/\a_3^2$ is the low-energy velocity of
the khronon. Upon proper translation of notations, this coincides with the
Lagrangian for the isocurvature mode obtained in~\cite{APSG}. From
(\ref{varphisup4}) we see that the isocurvature mode evolves slowly
with the rate suppressed by the slow-roll parameter $\ve_1$. This
behavior is also illustrated in Fig.~\ref{Fig:khronon}. 

Even if the isocurvature mode does not develop instability at late
times, it initially grows due to anti-friction, see
Eq.~(\ref{varphisup1}) and Fig.~\ref{Fig:khronon}. By the time the
growth terminates the power spectrum of $\cR$ reaches
\begin{equation}
{\cal P}_\cR\Big|_{a= p/M_*}=\frac{H^2}{4\pi^2m_{k,z}M_*}
\simeq \frac{H}{4\pi^2 M_P}\;.
\end{equation} 
For the validity of the linearized theory developed above we must
require that the perturbations of $\cR$ do not exceed unity. Then we
obtain an upper bound on the inflationary Hubble scale,
\begin{equation}
H<4\pi^2 M_P\;.
\label{Hbound}
\end{equation}

This constraint is somewhat unexpected, as a priori HL gravity should
be applicable also at trans-Planckian energies\footnote{Recall, in
  particular, that for $z=3$ the power spectra of the curvature
  perturbation $\zeta$ and the gravitational waves $\g$ do not depend
  on the Hubble scale, so their perturbative calculation does not
  require sub-Planckian energies.}.
In fact, the requirement (\ref{Hbound}) may be too restrictive. It
follows from consideration of metric perturbations
with very long wavelengths. Unlike
in GR, we cannot use a space-dependent reparameterization of
time to remove this perturbation completely. However, {\it
  space-independent} time reparameterizations are still a symmetry of
HL gravity and can be used to remove the fluctuation $\cR$ at any
given point. This suggests that coupling of khronon to other physical
degrees of freedom should involve spatial derivatives and its almost
homogeneous fluctuation, even with a large amplitude, should
not have any effect locally. This property is indeed satisfied by the
Lagrangian (\ref{Lagzeta2}) describing dynamics of $\zeta$ and khronon
at super Hubble scales. We have also verified
that in pure de Sitter universe the growth of $\cR$ does not lead to
divergence of local gauge invariant observables constructed out of the
metric $h_{ij}$, such as    
the extrinsic curvature $K$ and 
$K^i\!_j K^j\!_i$ and the spatial Ricci scalar $R$. 
For instance, the linear perturbation of the
trace part of the extrinsic curvature is given by 
\begin{align}
 & \delta K = K - 3H = 
-\frac{1- 2 \bar{\alpha}}{\bar\a} \left[ (1- \Omega_1)
 \dot{\cR} - \Omega_2 H \cR \right] \propto a^{-(z+ \frac{3}{2})}\,.
\end{align}
These arguments indicate that the constraint (\ref{Hbound}) may be
avoided by a more careful treatment where the growth of super Hubble
khronon fluctuations is absorbed by an appropriate field
redefinition. This study is, however, beyond the scope of the present
paper.

Before concluding this section, let us describe what happens if the
slow-roll parameter $\ve_1$ is the smallest quantity in the setup,
\begin{equation}
\ve_1/\a_1\ll 1\;.
\label{epsalhier1}
\end{equation}
In this case the perturbations $\cR$ and $\varphi$ are decoupled all
the way through the Hubble crossing down to $X \simeq \ve_1/\alpha$. After that, the good
variables are $\zeta$ and $\varphi$. As before, $\zeta$ is conserved,
whereas the evolution of $\varphi$ is described by
Eqs.~(\ref{varphisup1}), (\ref{varphisup2}), (\ref{varphisup3}),
(\ref{varphisup4}). The power spectrum of $\zeta$ is determined by
matching it to the fluctuations of the inflaton and khronon at $X \simeq \ve_1/\alpha$. It is easy to see that the inflaton fluctuations
dominate, so the spectrum is still given by
(\ref{Exp:powerzeta0}) (leaving aside a small correction due to the damping of the inflaton perturbations between the Hubble crossing and $X \simeq \ve_1/\alpha$). Note that for $z=3$ the hierarchy
(\ref{epsalhier1}) is actually not viable, as it would imply that the
power spectrum is larger than unity, see Eq.~(\ref{Pzeta3}).

\section{Violation of consistency relation}
\label{Sec:LV}
In the previous section, we computed the power spectra of the adiabatic curvature
perturbation $\zeta$ and the primordial gravitational waves in the anisotropic
scaling regime of HL 
gravity. In particular, we have shown that in the non-projectable case
$\zeta$ is conserved at super Hubble scales during inflation, despite
the presence of an isocurvature scalar perturbation. The
intuitive explanation of this conservation is that the isocurvature
mode, associated to the shift of khronon, is locally unobservable and
its interaction with $\zeta$ is suppressed by spatial
derivatives. This suggests that $\zeta$ will not be affected by the
isocurvature mode at super Hubble scales also after the end of
inflation. Indeed, conservation of $\zeta$ at super Hubble scales has
been demonstrated for rather general matter content in the low-energy
limit ($z=1$) of non-projectable HL gravity in
Refs.~\cite{APSG,Kobayashi:2009hh}. We will proceed under the
assumption that this also holds between the end of inflation and the
time when the universe enters into the isotropic scaling regime. 
Below we discuss a signal of the Lifshitz scaling in the
primordial spectra.

\subsection{Consistency relation in 4D Diff invariant  theories}
\label{ssec:crLI}
Before discussing the primordial spectra generated in the anisotropic
scaling regime, let us review the discussion in
theories encompassed by the Effective Field Theory (EFT) of inflation
\cite{Cheung:2007st} where the inflaton background breaks 
4D Diff invariance down to
time-dependent spatial Diff. We follow Ref.~\cite{CGNV}. 
Within EFT of inflation, the quadratic action for the gravitational waves is
 given by   
\begin{align}
S_{\gamma\gamma} = \frac{1}{8}\int{\dd t\, \dd^3 \bm{x} \, a^2\,
 \frac{M^2_P}{c_\g^2} \left[(\gamma'_{ij})^2 -
 c^2_\g (\partial_k{\gamma_{ij}})^2 \right]}\,.\label{EFTact}
\end{align}
In the presence of a time-dependent inflaton background
which breaks Lorentz invariance and time-translations, the parameters
$M_P$ and $c_\g$ can
deviate from their vacuum values and can vary with
time. However, one can always set these parameters to fixed values
by a redefinition of the metric.
Indeed, performing the
disformal transformation:
\begin{align}
g_{\mu\nu} \mapsto g_{\mu\nu} + (1 - c^2_\g(t))n_\mu n_\nu\,,
\end{align}
where $n_\mu$ is the unit vector orthogonal to the constant-inflaton slices,
and successively performing the conformal transformation to the
Einstein frame,
\begin{align}
g_{\mu\nu} \mapsto c^{-1}_\g(t)\frac{M_P^2(t)}{M^2_{P,0}}g_{\mu\nu}\,,
\end{align}
we can set the graviton speed $c_\g$ to unity and $M_P^2$ to
constant. 
The equivalence between the Einstein frame and
the Jordan frame for the gravitational waves was explicitly confirmed in 
Ref.~\cite{CGNV}. The price to pay is that these transformations
also alter the sector of scalar perturbations. For instance, if the
propagation
speed of the inflaton $c_s$ is 1 in the original frame, after the
above disformal transformation 
which sets $c_\g$ to 1, the sound speed $c_s$ is changed into 
$c_s = c^{-1}_\g$. 

After inflation, the non-minimal coupling introduced by the inflaton
should disappear.  
Therefore, it is reasonable to calculate the primordial
spectra in the Einstein frame for the gravitational waves. Then the
spectrum for the gravitational waves is given by the standard
expression 
and depends only on the ratio of inflationary Hubble scale and Planck
mass. Besides, one obtains the well-known 
consistency relation
\begin{align}
n_t  = -\frac{r}{8c_s}\,,\label{n_t}
\end{align}
which relates the spectral index for the gravitational waves $n_t$ and
the tensor to scalar ratio $r$. (The sub leading contribution to the
consistency relation in the slow-roll approximation can be found, e.g.,
in Ref.~\cite{Baumann:2014cja}.) In a Lorentz invariant theory the
velocity of any excitation cannot exceed unity, $c_s\leq 1$, which
implies a bound,
\begin{align}
\label{ntbound}
 & - n_t  \geq \frac{r}{8}\,.
\end{align}
This is a robust prediction of (single field) EFT of
inflation. Moreover, when $c_s$ is smaller than $1$ the equilateral
non-Gaussianity is enhanced by 
$1/c_s^2$ (see, e.g., Refs.~\cite{Seery:2005wm, Baumann:2011su,
Ade:2015ava}). Thus, a deviation from equality in (\ref{ntbound})
should be accompanied by large non-Gaussianity.

\subsection{Violation of consistency relation in Ho\v rava--Lifshitz gravity}
\label{ssec:crLV}
We now discuss the primordial spectra generated in gravity with anisotropic
scaling. In this case the symmetry breaking pattern is different:
there are no 4D Diff to start with, but only the reduced
symmetry of foliation-preserving Diff, that is further broken to
time-dependent spatial Diff by the inflaton background.
The velocity of graviton depends now on 
the wavenumber $p$, so one cannot
set it to unity by the disformal transformation which
globally changes the time component of the metric.
This means that the modified dispersion relation
physically changes the spectrum of the gravitational waves. 
In particular, the relation between the power spectrum ${\cal P}_\g$,
Eq.~(\ref{Exp:powertens}), and the inflationary Hubble rate
is no longer straightforward: it depends on the scaling exponent $z$
and other parameters of the theory. For $z=3$ the tensor power
spectrum does not depend on $H$ at all. On the other hand, a robust
prediction for $z=3$ is vanishing of the tensor spectral index, $n_t=0$.

Using Eqs.~(\ref{Exp:powertens}) and
(\ref{Exp:powerzeta0}), at the leading order in the slow-roll
approximation, we obtain the tensor-to-scalar ratio as\footnote{
Equations (\ref{Exp:powertens}) and (\ref{Exp:powerzeta0}) directly give
\begin{align}
 & r = 16 \varepsilon_1
 \left( \frac{\vk_z}{\vk_{\gamma, z}} \right)^{\frac{3}{2z}}
 \left( \frac{H_{p, \gamma}}{H_p} \right)^{\frac{3}{z}-1} \,. \label{Exp:r}
\end{align}
For $\vk_z \neq \vk_{\gamma, z}$, the Hubble crossing times for the
adiabatic perturbation does not necessarily coincide with the one for
the gravitational waves and the Hubble parameters at these times are
related as
\begin{align}
 & \frac{H_{p, \gamma}}{H_p}  \simeq  \left(
 \frac{\vk_z}{\vk_{\gamma,\,z}} \right)^{\frac{\varepsilon_1}{2z}}\,.
\end{align}
} 
\begin{align}
 & r \equiv \frac{{\cal P}_\gamma}{{\cal P}_\zeta} \simeq 16 \varepsilon_1
 \left( \frac{\vk_z}{\vk_{\gamma, z}} \right)^{\frac{3}{2z}} \,. \label{Exp:rSR}
\end{align}
Exceptionally, for $\vk_z = \vk_{\gamma, z}$ the tensor-to-scalar ratio is given by the standard expression irrespective of the
value of $z$. 
Using Eqs.~(\ref{Exp:nt}) and (\ref{Exp:rSR}), we obtain the modified
consistency relation for the primordial perturbations in the anisotropic scaling regime as
\begin{align}
 & n_t \simeq - \frac{3- z}{z} \frac{r}{16} \left(
 \frac{\vk_{\gamma, z}}{\vk_z} \right)^{\frac{3}{2z}}  \,. \label{HL_nt}
\end{align}
We see that $n_t$ and $r$ are still related linearly, but the
coefficient depends on 
$z$, $\vk_z$, and $\vk_{\gamma, z}$. 
Clearly, this can violate the lower bound (\ref{ntbound}) on
$-n_t$ obtained in Lorentz invariant theories.


\section{Concluding remarks}
\label{Sec:concl}
HL gravity contains an additional scalar
degree of freedom in the gravity sector, khronon, corresponding to
fluctuations of the preferred
time foliation. Therefore, a minimal model of inflation possesses two
scalar degrees of freedom: the 
inflaton and khronon. These two fields are coupled gravitationally. 
In the small scale limit, as usual,
the gravitational interaction is suppressed and we simply have two
decoupled Lifshitz scalar fields. Naively, one may expect that in the
large scale limit, the gravitational interaction becomes important and
these two fields start to be coupled. This is indeed the case in
the projectable version of HL gravity. The inflaton and khronon stay nearly
gapless modes which are bi-linearly coupled. Then the adiabatic curvature
perturbation $\zeta$ is generically not conserved at large scales. 

On the other hand, the situation is crucially different in the
non-projectable version. In the anisotropic scaling regime, khronon acquires the effective mass $m_K$, which is much
larger than the Hubble scale, well before 
Hubble crossing time. It then decouples from the adiabatic
mode $\zeta$ and does not leave any impact on 
the power spectrum of $\zeta$, which is conserved at super Hubble scales. The power spectrum of $\zeta$ is simply
given by that of the Lifshitz scalar with the multiplicative factor 
$1/(2\varepsilon_1 M_P^2)$. The decoupled khronon rapidly oscillates,
with the amplitude of the oscillations growing 
exponentially due to anti-friction. The growth persists until the mode
enters into the regime of isotropic scaling as a consequence of the
redshift of its momentum.
We need a
more careful consideration to see if this exponential growth can or
cannot 
affect
observable quantities.

One remaining question is whether the decoupling between the adiabatic mode
$\zeta$ and khronon is a robust feature of
non-projectable HL gravity also beyond the restricted setup considered
in this paper.
We have focused on
the linear order in perturbations. The physical
interpretation presented in Sec.~\ref{ssec:NP} suggests that the decoupling
will also persist at non-linear orders. We
postpone an explicit analysis of this issue, as well as of primordial
non-Gaussianity, to a future work. In this
paper we assumed the minimal coupling of the inflaton to the gravity
sector. One may wonder whether a non-minimal interaction can prevent the 
decoupling of khronon. Recall that khronon gets gapped due to a
peculiar structure of the
coefficient in front of the (quadratic) time
derivative term in the action. Thus, to make khronon gapless, the
non-minimal coupling should modify the 
time derivative terms. The only contribution that can
change the time derivative terms under the assumption of 
foliation-preserving Diff and time reversal symmetry is the term with
$K \dot{\Phi}/N$. However, this can be removed by a redefinition of
the metric $h_{ij} \to \Omega^2(\Phi) h_{ij}$ and 
$N \to \Omega^3(\Phi) N$. Therefore, we expect that the decoupling
between $\zeta$ and khronon takes place generically in the
non-projectable version of HL gravity with the time reversal symmetry in
the anisotropic scaling regime. It may be interesting to study if this
decoupling takes place also in the case when the time reversal symmetry is broken,
e.g. by a term with $\sqrt{h}\, \dot{\Phi}/N$. 

We also pointed out that the consistency relation between the tensor to
scalar ratio $r$ and the tensor spectral index $n_t$, which holds in the
general single field EFT of inflation, can be violated by the primordial
perturbations generated during the anisotropic scaling regime. If the
primordial gravitational waves are detected, the value of $r$ will give the
lower bound on $- n_t$ in Lorentz invariant theory. A violation of
this bound will indicate violation of Lorenz invariance in the early universe.

\acknowledgments
Y.~U. would like to thank Jaume Garriga for his fruitful
comments. Y.~U. would like to thank CERN for the hospitality
during the work on this project. S.~A. is supported by 
Japan Society for the Promotion of Science (JSPS) under Contract
No.~17J04978 and in part by  
by Grant-in-Aid for Scientific Research on Innovative Areas under Contract
No.~15H05890. S.~S. is supported by the RFBR grant No.~17-02-00651.
Y.~U. is supported by JSPS Grant-in-Aid for Research Activity Start-up
under Contract No.~26887018 
and Grant-in-Aid for Scientific Research on Innovative Areas under
Contract No.~16H01095. Y.~U. is also supported in part by Building of
Consortia for the Development of Human Resources in 
Science and Technology, Daiko Foundation, and the National Science Foundation under Grant No. NSF
PHY11-25915.

\appendix
\section{Quadratic action} 
 \label{app:A}

In this Appendix we give the complete expression of the quadratic action
for the mixed system of the inflaton and khronon.

\subsection{Action for $\cR$ and $\varphi$}
Substituting the fields (\ref{Exp:ADMmetric}) into the Lagrangian
consisting of (\ref{Lagrfull}) and (\ref{LagrInf}), expanding to
second order in perturbations and integrating out
the lapse function and
the shift vector, we obtain
\begin{align}
 & S= \int \dd^4 x {\cal L} 
  = \int \dd t \int \dd^3 \bm{p} \left[ {\cal L}_{\cR} +
 {\cal L}_{\varphi} + {\cal L}_{\cR \varphi} \right]
\end{align}
with
\begin{align}
 & {\cal L}_{\cR}=a^2  M_*^2 \frac{1+
 \bar{\alpha}}{\alpha_1\bar\a}  
\left[(1- \Omega_1(t,\, p))\cR'_{\sbm{p}}
 \cR'_{- \sbm{p}} - \omega_{\cR}^2(t,\,p) \cR_{\sbm{p}}
 \cR_{- \sbm{p}} \right]\,, \\
 & {\cal L}_{\varphi}= \frac{a^2}{2} \left[ \left(1
 - \frac{\bar{\alpha} \varepsilon_1 }{1- 2 \bar{\alpha}}  \Omega_1(t,\, p)
 \right) \varphi'_{\sbm{p}} \varphi'_{- \sbm{p}}  -
 \omega^2_{\varphi}(\eta,\,p)  \varphi_{\sbm{p}}  \varphi_{-\sbm{p}} \right]
 \,, \\
 & {\cal L}_{\cR \varphi}= a^2\bigg\{  -
 \frac{\phi'}{\cal H} \Omega_1(t,\, p)  \varphi'_{\sbm{p}} \cR'_{- \sbm{p}} +
 \left[  \frac{1- 2 \bar{\alpha}}{\bar{\alpha}} (1- \Omega_1(t,\, p))
 \phi' + \Omega_1(t,\,p) \frac{\phi'' - {\cal H} \phi'}{{\cal H}} \right]
 \varphi_{\sbm{p}} \cR'_{- \sbm{p}}  \nonumber \\
 & \qquad \qquad \qquad \qquad  - \phi' \Omega_2(t,\, p) 
 \varphi'_{\sbm{p}} \cR_{-\sbm{p}} - \left( \frac{1+ \bar{\alpha}}{\bar{\alpha}}
  {\cal H} \phi'  + a^2 V_{\phi} \right) \Omega_2(t,\,p) \varphi_{\sbm{p}}
 \cR_{- \sbm{p}}\bigg\}  \,,\label{mixLag}
\end{align}
where the functions $\Omega_{1,2}(t,p)$ have been introduced in
(\ref{Def:Omega1}), (\ref{Def:Omega2}), the
frequency $\omega_{\cR}$ is given by Eq.~(\ref{Exp:omegaRMB})
and $\omega_{\varphi}$ is given by 
\begin{align}
 &  \frac{\omega^2_{\varphi}(t,\,p)}{{\cal H}^2}=
 \biggl(\frac{p}{{\cal H}}\biggr)^2 \left[ \vk_1 + \vk_2
  \biggl( \frac{p}{a M_*} \biggr)^2  + \vk_3
  \biggl( \frac{p}{a M_*} \biggr)^4  \right] + \frac{a^2 V_{\phi\phi}}{{\cal H}^2} -
 \frac{1+\bar\a}{\bar\a}\varepsilon_1 \cr
 & \qquad \qquad \qquad \qquad \quad + \frac{(1+
   \bar{\alpha})\alpha_1}{2 \bar{\alpha} M_*^2} 
 \biggl( \phi' + \frac{\bar{\alpha}}{1+
 \bar{\alpha}} \frac{a^2 V_\phi}{\cal H} \biggr)^{\!2} 
\frac{\Omega_1(t,\,p)}{{\cal H}^2}
  \cr 
 & \qquad \qquad \qquad \qquad \quad - \frac{\alpha_1}{2
 M_*^2}  \frac{1}{({\cal H} a)^2} \biggl[a^2 \phi'\bigg( \phi'
 +  \frac{\bar{\alpha}}{1+ \bar{\alpha}}\frac{a^2V_\phi}{\cal H}\bigg)
 \frac{\Omega_1(t,\,p)}{\cal H} \! \biggr]'. 
\end{align}
By inspection of various terms in the Lagrangian we can see that 
$\cR$ and $\varphi$
are decoupled in the limit of large momenta $p$.

\subsection{Action for $\zeta$ and $\varphi$}
Using $\zeta$ defined in (\ref{Def:zeta}) 
and eliminating $\cR$, we obtain the quadratic action as
\begin{align}
 & S  = \int \dd t \int \dd^3 \bm{p} \left[ {\cal L}_{\zeta} +
 \tilde{{\cal L}}_{\varphi} + {\cal L}_{\zeta \varphi} \right]
\end{align}
with
\begin{align}
 & {\cal L}_\zeta=a^2 M_*^2 \frac{1+
 \bar{\alpha}}{\alpha_1\bar\a} \left[(1- \Omega_1(t,\, p))  \zeta'_{\sbm{p}}
 \zeta'_{- \sbm{p}} - \omega_{\cR}^2(t,\,p)  \zeta_{\sbm{p}}
 \zeta_{- \sbm{p}} \right] \,,  \label{Lagzeta1} \\
& \tilde{\cal L}_{\varphi} = \frac{a^2}{2} \left[ \frac{(1-
 2\bar{\alpha}) \alpha_1}{2(1+\bar{\alpha})\varepsilon_1} \left( 1 +
 \frac{\bar{\alpha} \varepsilon_1}{1-2\bar{\alpha}} \right)
 \Omega_3(t,\,p) \varphi'_{\sbm{p}} \varphi'_{- \sbm{p}} -
 \tilde{\omega}^2_{\varphi}(t,\,p)  \varphi_{\sbm{p}} \varphi_{- \sbm{p}}
 \right]\,,  \label{tL}\\
 & {\cal L}_{\zeta \varphi} =a^2\bigg\{ M_*^2
 \frac{{\cal H}}{\phi'} \Omega_3(t,\,p)
 \left(  \zeta'_{\sbm{p}} \varphi'_{-\sbm{p}} - \frac{\phi'' - {\cal
       H} \phi'}{\phi'} 
 \zeta'_{\sbm{p}} \varphi_{-\sbm{p}} \right) -  \phi'
\Omega_2(t,\,p) \zeta_{\sbm{p}} 
 \varphi'_{-\sbm{p}} \cr
 &\quad \qquad \quad \quad - \left[ 2 M_*^2 \frac{1+ \bar{\alpha}}{\alpha_1\bar\a}
 \frac{{\cal H}}{\phi'} \omega^2_{\cR}(t,\,p)  + \! \left( \frac{1+
 \bar{\alpha}}{\bar{\alpha}} {\cal H} \phi'  + a^2 V_\phi \right)
 \Omega_2(t,\,p) \right]\! \zeta_{\sbm{p}} \varphi_{-\sbm{p}}\bigg\} \,,
 \label{Lagzeta2} 
\end{align}
where we introduced
\begin{align}
 & \Omega_3(t,\, p) = \biggl(\frac{p}{{\cal H}} \biggr)^2
 \biggl[1+ \frac{1}{\beta_4} \biggl( \frac{p}{a M_*} \biggr)^2  +
 \frac{1}{\gamma_4}  \biggl( \frac{p}{a M_*} \biggr)^4 \biggr] \Omega_1(t,\,p)\,.
\end{align}
The new expression for the $\varphi$-frequency is
\begin{align}
 & \frac{\tilde{\omega}_{\varphi}^2 (t,\,p)}{{\cal H}^2}
=\biggl(\frac{p}{{\cal H}}\biggr)^2 \left[ \vk_1 + \vk_2
  \biggl( \frac{p}{a M_*} \biggr)^2  + \vk_3
  \biggl( \frac{p}{a M_*} \biggr)^4  \right] + \frac{1-2
  \bar{\alpha}}{ \varepsilon_1\bar\a} 
 \frac{\omega_{\cR}^2(t,\,p)}{{\cal H}^2} \cr
 & \qquad \qquad \quad - \frac{\alpha_1 \bar{\alpha}}{2(1+
   \bar{\alpha})} \left\{ \frac{1- 2 
 \bar{\alpha}}{\bar{\alpha}} \varepsilon_1 + \left( \varepsilon_1 -
 \frac{\varepsilon_2}{2}\right)^2 \right\} \Omega_3(t,\,p) \cr
 & \qquad \qquad \quad + \frac{1}{(a {\cal H})^2} \left\{
 a^2 {\cal H} \left[ - \Omega_2(t,\,p) + \frac{(1- 2
 \bar{\alpha})\alpha_1}{2(1+\bar{\alpha})\varepsilon_1} \left( \varepsilon_1
 - \frac{\varepsilon_2}{2} \right)
\bigg(1+\frac{\varepsilon_1\bar\a}{1-2\bar\a}\bigg)
 \Omega_3(t,\,p)  \right] \right\}' \cr
 & \qquad \qquad \quad + 2 \left( \frac{1-2\bar{\alpha}}{\bar{\alpha}} +
 \varepsilon_1 - \frac{\varepsilon_2}{2} \right) \Omega_2(t,\, p) -
\frac{(1-2\bar{\alpha}) 
 \alpha_1}{8(1+ \bar{\alpha})} \frac{\varepsilon_2}{\varepsilon_1}
 (\varepsilon_2- 4 \varepsilon_1) \Omega_3(t,\,p). \cr 
\label{tomegaphi}
\end{align}
Notice that all terms in ${\cal L}_{\zeta \varphi}$ and 
$\tilde{\cal L}_{\varphi}$ are multiplied by factors of order ${\cal O}(X)$. 
This implies that $\zeta$ has a constant solution in the long-wavelength
limit.
As discussed in the main text,
the degree of freedom which is orthogonal to $\zeta$ acquires a
mass gap which is much larger than $H$ in the anisotropic scaling regime.

\section{Khronon-inflaton mixing for $z=1$}
\label{app:iso}

If the inflationary Hubble rate is low, $H<M_*\sqrt{\a_1/\ve_1}$,
mixing between the inflaton and khronon perturbations 
occurs in the regime where the
dynamics is dominated by the terms with relativistic scaling
$z=1$. In this Appendix, we consider the case with 
$\varepsilon_1/\alpha> 1$ and the range $X \ll 1/\alpha^2$. 

Compared to the Lagrangian (\ref{Lagmixed}) considered in
the main text, one should keep an additional mixing contribution, so
that the total mixing Lagrangian reads,
\begin{equation}
{\cal L}_{\cR\varphi}=a^2\,\bigg[
-\frac{\varphi_{\sbm{p}}'\hat
  R_{-\sbm{p}}'}{\sqrt{1+\frac{\a_1X}{2\ve_1}}}
+\frac{\a_1}{\a_3}\frac{{\cal H}^2X}{\sqrt{1+\frac{\a_1X}{2\ve_1}}}
\varphi_{\sbm{p}}\hat R_{-\sbm{p}}\bigg]\;.
\end{equation} 
The quantity $X$ is now given simply by,
\[
X=(p/{\cal H})^2\;,
\]
whereas the fields' frequencies are,
\[
\hat\omega_\cR^2=p^2\frac{\a_1}{\a_3}
\frac{1+\frac{\a_1\bar\a}{\a_3\ve_1}X}{1+\frac{\a_1X}{2\ve_1}}
~,~~~~~~
\omega_\varphi^2=\vk_1p^2\;.
\] 
At $X\gg \ve_1/\a_1$ the fields $\varphi$ and $\hat \cR$ are decoupled
and have the same velocities as the inflaton and khronon in flat
spacetime, whereas at $X\ll \ve_1/\a_1$ they become strongly mixed. To
diagonalize the Lagrangian in the latter case, we write it in terms of
$\varphi$ and 
\[
\chi_+=\varphi-\frac{\hat\cR}{\sqrt{1+\frac{\a_1X}{2\ve_1}}}\;.
\]
It is straightforward to see that the mixing terms are negligible
at $X\ll \ve_1/\a_1$. Thus, we conclude that in this regime the
decoupled modes are $\chi_+$ (or equivalently $\zeta$, see
Eq.~(\ref{Def:zeta})) and $\varphi$. Their Lagrangian reads,
\begin{equation}
\label{Lagz1}
{\cal L}=\frac{a^2}{2}
\Big({\chi_+'}^2-\frac{\a_1}{\a_3}p^2\,\chi_+^2\Big)
+\frac{a^2 \a_1}{4\ve_1}\bigg(\frac{p}{\cal H}\bigg)\bigg[
\varphi'^2-{\cal
  H}^2\frac{2\ve_1}{\a_1}\Big(\vk_1-\frac{\a_1}{\a_3}\Big)
\,\varphi^2\bigg]\;.
\end{equation}
It is worth stressing that in this Appendix we have focused on
`sub Hubble' modes, i.e. modes with $X\gg 1$. Nevertheless, we
observe that the $\varphi$-equation following from the Lagrangian
(\ref{Lagz1}) coincides with Eq.~(\ref{varphisup3}) obeyed by
super Hubble isocurvature modes in the $z=1$ regime. In other words,
in the $z=1$ case the adiabatic and isocurvature modes are described
respectively by $\zeta$ and $\varphi$ at all times when $X\ll
\ve_1/\a_1$. Note that the velocity of the adiabatic mode is given by
$\sqrt{\a_1/\a_3}$ and coincides with the velocity of gravitons,
rather than the velocity of inflaton.

\end{document}